%% file: main.tex
\documentclass[
reprint,
amsmath,amssymb,
superscriptaddress,
pra,
]{revtex4-2}

\usepackage{graphicx}
\usepackage{dcolumn}
\usepackage{bm}
\usepackage{physics}
\usepackage{braket}
\usepackage{placeins}
\usepackage[colorlinks=true]{hyperref}
\usepackage{upgreek}
\usepackage{multirow}
\usepackage{dsfont}
\usepackage{color,soul}
\begin{document}

\let\oldaddcontentsline\addcontentsline
\renewcommand{\addcontentsline}[3]{}
\newcommand{\citesec}[2]{\cite[\S{}#2]{#1}}

\input{main_header.tex}

\date{\today}

\maketitle

\input{manuscript.tex}
\bibliography{library}

\FloatBarrier
\clearpage
\onecolumngrid
\begin{center}
{\Large \textbf{Supplementary information}}
\end{center}
\makeatletter
   \renewcommand\l@section{\@dottedtocline{2}{1.5em}{2em}}
   \renewcommand\l@subsection{\@dottedtocline{2}{3.5em}{2em}}
   \renewcommand\l@subsubsection{\@dottedtocline{2}{5.5em}{2em}}
\makeatother
\let\addcontentsline\oldaddcontentsline

\renewcommand{\thesection}{\arabic{section}}


\let\oldaddcontentsline\addcontentsline
\renewcommand{\addcontentsline}[3]{}
\let\addcontentsline\oldaddcontentsline
\renewcommand{\theequation}{S\arabic{equation}}
\renewcommand{\thefigure}{S\arabic{figure}}
\renewcommand{\thetable}{S\arabic{table}}
\renewcommand{\thesection}{S\arabic{section}}
\setcounter{figure}{0}
\setcounter{equation}{0}
\setcounter{section}{0}

\newcolumntype{C}[1]{>{\centering\arraybackslash}p{#1}}
\newcolumntype{L}[1]{>{\raggedright\arraybackslash}p{#1}}

\FloatBarrier

\input{SI.tex}

\end{document}

%% file: main_header.tex
\title{
    Single-qubit gates with errors at the $10^{-7}$ level\\
}

\begin{abstract}
    We report the achievement of single-qubit gates with sub-part-per-million error rates, in a trapped-ion $^{43}$Ca$^{+}$ hyperfine clock qubit.
    We explore the speed/fidelity trade-off for gate times $4.4\leq t_{g}\leq35~\upmu$s, and benchmark a minimum error per Clifford gate of $1.5(4) \times 10^{-7}$.
    Calibration errors are suppressed to $< 10^{-8}$, leaving qubit decoherence ($T_{2}\approx 70$ s), leakage, and measurement as the dominant error contributions.
    The ion is held above a microfabricated surface-electrode trap which incorporates a chip-integrated microwave resonator for electronic qubit control; the trap is operated at room temperature without magnetic shielding.
 \end{abstract}

\author{M.\,C.\,Smith}
\thanks{These authors contributed equally to this work.}
\affiliation{Clarendon Laboratory, Department of Physics, University of Oxford, Parks Road, Oxford OX1 3PU, U.K.}

\author{A.\,D.\,Leu}
\thanks{These authors contributed equally to this work.}
\affiliation{Clarendon Laboratory, Department of Physics, University of Oxford, Parks Road, Oxford OX1 3PU, U.K.}

\author{K.\,Miyanishi}
\affiliation{Clarendon Laboratory, Department of Physics, University of Oxford, Parks Road, Oxford OX1 3PU, U.K.}
\affiliation{Center for Quantum Information and Quantum Biology (QIQB),\\
The University of Osaka, 1-2 Machikaneyama, Toyonaka 560-0043, Japan}

\author{M.\,F.\,Gely}
\affiliation{Clarendon Laboratory, Department of Physics, University of Oxford, Parks Road, Oxford OX1 3PU, U.K.}

\author{D.\,M.\,Lucas}
\affiliation{Clarendon Laboratory, Department of Physics, University of Oxford, Parks Road, Oxford OX1 3PU, U.K.}

%% file: manuscript.tex
\begin{figure}[h!]
    \centering
    \includegraphics[width=0.45\textwidth]{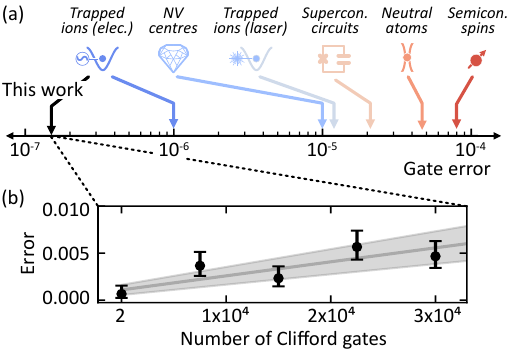}
    \caption{
    \textbf{Single-qubit gates: state-of-the-art.}
    \textbf{(a)} State-of-the-art single-qubit gates across different quantum computing platforms. From left to right: electronically driven trapped ions~\cite{harty2014}, NV centres~\cite{Bartling2024}, laser-driven trapped ions~\cite{moses2023}, superconducting circuits~\cite{rower2024}, neutral atoms~\cite{Sheng2018,nikolov2023}, and semiconductor spin qubits~\cite{lawrie2023}.
    \textbf{(b)} Single-qubit gate randomised benchmarking (RB) measurement carried out in this work.
    The average gate duration is $13\ \upmu$s; the fit yields an average Clifford gate error of $1.5(4) \times 10^{-7}$ and an average state-preparation and measurement (SPAM) error of $1.1(5)\times 10^{-3}$, which is consistent with independent SPAM measurements.
    }
    \label{fig:fig1}
\end{figure}

High-fidelity quantum logic operations are essential for realising fault-tolerant quantum computation~\cite{divincenzo2000}, as lowering gate errors considerably reduces the number of physical qubits and their control circuitry required for error correction~\cite{fowler2012}.
In this respect, trapped ion qubits are one of the most promising approaches to quantum information processing, and have been used to demonstrate gate fidelities~\cite{loschnauer2024,leu2023,duwe2022,benhelm2008,clark2021,srinivas2021,zarantonello2019,hahn2019,schafer2018,gaebler2016,weidt2016,ballance2016,harty2016,harty2014} and state-preparation and measurement~\cite{sotirova2024,an2022,ransford2021,harty2014} well beyond error correction thresholds.
A major advantage of trapped ions is the existence of ``atomic clock'' transitions for encoding qubits.
These are transitions which, in the presence of the correct static magnetic field, are insensitive to changes in magnetic field (to first order), enabling coherence times in excess of an hour~\cite{wang2021}.
The gigahertz frequency of hyperfine clock transitions also enables qubit state manipulation using low-cost electronic sources with exceptionally stable amplitude and phase (compared with laser sources).
Electronically-controlled trapped ions have been used to perform the highest fidelity single- and two-qubit operations of any platform, with gate errors of $1.0(3) \times 10^{-6}$ and $3(1) \times 10^{-4}$ respectively~\cite{harty2014,loschnauer2024}.
Even though single-qubit gates typically have higher fidelities than their two-qubit counterparts, quantum algorithms often require many more single-qubit than multi-qubit operations, and are therefore more sensitive to single-qubit gate errors~\cite{NielsenChuang2010}.
High-fidelity single-qubit operations also have many other uses both in quantum information and further afield: for protecting ``idle'' qubits through dynamical decoupling~\cite{viola1999}; in composite pulse sequences for error compensation~\cite{levitt1986} or addressing individual qubits~\cite{leu2023,loschnauer2024}; and in quantum sensing applications~\cite{kotler2013,soare2014}.
In this Letter, we demonstrate microwave-controlled single-qubit gates, using a trapped-ion clock qubit, with errors benchmarked at $1.5(4) \times 10^{-7}$.
This result constitutes nearly an order of magnitude improvement in both fidelity and gate speed over the previous best result \cite{harty2014}, see Fig.~\ref{fig:fig1}(a).
We describe the calibration and error characterisation methods that led to this result, which are readily applicable to all types of physical qubits.
We give a comprehensive breakdown of known sources of error; our model predicts a gate error of $1.7(1) \times 10^{-7}$.
Given the technical nature of the dominant errors, we thereby chart a path to even higher fidelity operations.

\begin{figure}[h]
    \centering
    \includegraphics[width=0.45\textwidth]{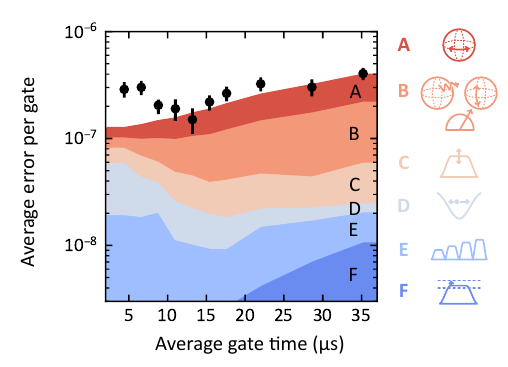}
    \caption{
    \textbf{Gate error contributions.}
    The data points show RB measurements for different gate durations.
    Each point collates data from six different measurement runs carried out on six different days.
    Shaded areas show the contributions from known sources of error: \textbf{A} decoherence; \textbf{B} leakage, bit-flip and time-dependent measurement errors; \textbf{C} fast amplitude noise; \textbf{D} harmonic motion; \textbf{E} slow amplitude drift; \textbf{F} amplitude resolution.
    }
    \label{fig:fig2}
\end{figure}

Experiments are carried out on a segmented surface-electrode Paul trap, used to confine an ion 40 $\upmu$m above the chip surface.
An on-chip microwave (MW) resonator is used, together with the MW drive chain detailed in~\citesec{SuppInfoLabel}{\ref{SI:MW}}, to generate a MW field for driving quantum logic operations.
Our qubit is defined by the hyperfine states $\ket{1} = \ket{F = 4, M = 1}$ and $\ket{0} = \ket{F = 3, M = 1}$ of the ground manifold $4^{2}S_{1/2}$ of $^{43}$Ca$^{+}$.
These states form a clock qubit at our static magnetic field strength of 28.8 mT and are connected by a magnetic dipole transition with a frequency splitting of $ 3.123$ GHz.
Further details about the trap design and choice of qubit can be found in Ref.~\cite{weber2022} (although it should be noted that the trap is operated at room temperature in the present work).
Gate errors are measured through randomised benchmarking (RB)~\cite{knill2008, kwiatkowski2023}.
The ion is randomly prepared in either state $\ket{0}$ or $\ket{1}$~\cite{harty2014,leu2024} before being subject to a pseudo-random series of Clifford gates.
Each Clifford gate is decomposed into $\pm \pi/2$ pulses in the $\hat{\sigma}_{x}$ and $\hat{\sigma}_{y}$ bases, with an average of 2.2 pulses per Clifford.
We then randomly shelve either $\ket{0}$ or $\ket{1}$ into the ``dark'' $3\text{D}_{5/2}$ manifold and measure the state of the ion using state-dependent fluorescence~\cite{myerson2008}.
For each of the 5 different sequence lengths employed, we use 30 random gate sequences -- with different sequences used for each gate time studied -- and we repeat 100 shots of each sequence.
The increase in error with sequence length yields an error of $1.5(4) \times 10^{-7}$ per Clifford gate for a gate duration of $13\ \upmu$s, as shown in Fig.~\ref{fig:fig1}(b).
Results for a range of gate durations -- between $4.4\ \upmu$s and $35\ \upmu$s -- are shown in Fig.~\ref{fig:fig2}.
Further information regarding the implementation and analysis of the RB measurements can be found in~\citesec{SuppInfoLabel}{\ref{SI:RBM}}.
By independently measuring or simulating all known sources of error, we predict a Clifford gate error of $1.7(1) \times 10^{-7}$ for a gate duration of $13\ \upmu$s.
Table~\ref{tab:error_budget} summarises the different error sources (see also~\citesec{SuppInfoLabel}{\ref{SI:Errors}}).
We find good agreement between the predicted and measured error, as shown in Fig.~\ref{fig:fig2}.
The prediction begins to deviate from the measured error at shorter gate times, which indicates the existence of further error sources that we have not taken into account.
We now discuss the different sources of error, giving specific values for the case of 13~$\upmu$s gates.
\begin{table}[t]
    \setlength{\tabcolsep}{12pt}
    \begin{tabular}{ll}
    Error source            & Error ($/ 10^{-7}$ )                 \\
    \hline\rule{0pt}{2.5ex}\textbf{A}. Decoherence                              & 0.64(7)     \\
    \textbf{B}. Leakage, bit-flip and  & \\
    $\ \ \ \ \ $time-dependent measurement                      & 0.62(7)    \\
    \textbf{C}. Fast amplitude noise                          & 0.23(2)    \\
    \textbf{D}. Harmonic motion                          & 0.13(2)    \\
    \textbf{E}. Slow amplitude drift                          & 0.09(7)     \\
    ac Zeeman shift calibration                 & 0.03(2)    \\
    \textbf{F}. Amplitude resolution                     & 0.015     \\
    Excitation of spectator states              & $<$0.01     \\
    Inter-pulse Zeeman shift ramping          & $<$0.01     \\
    Non-rotating-wave dynamics          & $<$0.001     \\
    \hline\hline\rule{0pt}{2.5ex}Expected error per gate                              & 1.7(1)   \\
    Measured error per gate                             & 1.5(4)
    \end{tabular}
    \caption{
    \textbf{Errors for a 13~$\upmu$s Clifford gate.}
    These estimates of the different error contributions are either calculated from simulations or measured independently of the RB measurements.
    Estimated uncertainties are given in parentheses.
    The errors displayed in Fig.~\ref{fig:fig2} use the same lettering.
    The ac Zeeman shift calibration error is not plotted in Fig.~\ref{fig:fig2}, as it is negligibly small for most gate times, and can be readily reduced with our calibration procedure illustrated in Fig.~\ref{fig:fig4}.
    }
    \label{tab:error_budget}
\end{table}

\begin{figure}[h]
    \centering
    \includegraphics[width=0.45\textwidth]{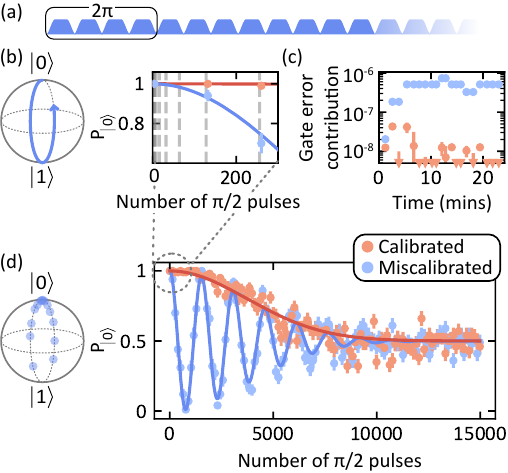}
    \caption{
    \textbf{Pulse amplitude calibration, drift, and fluctuations.}
    \textbf{(a)} Changes in amplitude are measured by driving multiples of four $\pi/2$ pulses, such that ideally the ion is returned to state $\ket{0}$ at the end of the sequence.
    \textbf{(b)} The number of pulses is increased exponentially and, when the probability $P_{\ket{0}}$ of recovering the initial state is below a chosen threshold, the amplitude miscalibration is computed and compensated for.
    \textbf{(c)} Calibrations interleaved in the RB measurement process reveal the gate error that would arise from drifts (blue), and the reduced error obtained through repeated calibration (orange).
    \textbf{(d)} The accuracy of this calibration procedure is limited by shot-to-shot amplitude fluctuations.
    Amplitude changes between sequential shots of the experiment randomise the final state, reducing the reproducibility of measuring state $\ket{0}$ at the end of the sequence, so that $P_{\ket{0}}\rightarrow 0.5$ for long sequences.
    }
    \label{fig:fig3}
\end{figure}

Pulses are generated by an arbitrary waveform generator (AWG) at $\sim100$ MHz, and are then upconverted to MW frequencies, amplified, and filtered, before being delivered to the on-chip resonator (see~\citesec{SuppInfoLabel}{\ref{SI:MW}}).
The pulse amplitude is ramped up and down over 40 ns to reduce off-resonant excitation to other states in the $4\text{S}_{1/2}$ ground manifold.
The error arising from the residual excitation to these states is estimated through simulation to be $<1 \times 10^{-9}$ per gate.
We use the minimum inter-pulse delay allowed by the AWG (40 ns) to minimise leakage, bit-flip and time-dependent measurement errors due to unnecessary dead time.
To calibrate the gates, we fix the pulse duration, then determine the optimal MW amplitude and frequency to drive a $\pi/2$ rotation.
For the amplitude calibration, we drive multiples of four $\pi/2$ pulses such that, in the ideal case, the qubit undergoes $2\pi$ rotations on the Bloch sphere as shown in Fig.~\ref{fig:fig3}(a).
An offset in the amplitude will result in a net rotation of the qubit state at the end of the sequence, from which we can infer the optimal pulse amplitude.
For efficiency, the number of $\pi/2$ pulses is increased exponentially until a statistically significant error signal has accrued, see Fig.~\ref{fig:fig3}(b).
In practice, we end the sequence with an additional $\pi/2$ pulse, such that a measurement of $P_{\ket{0}}=|\braket{\psi|0}|^{2}$, where $\ket{\psi}$ is the qubit state at the end of the sequence, reveals an amplitude offset as a deviation from $P_{\ket{0}}=1/2$.
This makes the calibration linearly sensitive to small offsets in amplitude and indicates the sign of the offset.
We interleave this amplitude calibration with RB measurements, running it approximately every minute (the calibration procedure takes a few seconds).
The calibration rate was chosen to measure the ``slow'' drift in amplitude on a timescale of minutes -- as shown in Fig.~\ref{fig:fig3}(c) -- rather than to maximise the RB run time, and could be run less frequently without compromising the gate error.
We suspect that a dominant cause of amplitude drift is heating of the amplifier in the MW drive chain, given the correlation observed between its logged temperature and the calibrated pulse amplitude.
These amplitude drifts were reduced by delivering a continuous MW drive, at the same amplitude and frequency used to drive gates, except during SPAM transfer pulses or RB sequences.
This keeps the MW power dissipated in the drive chain constant, improving temperature stability once a thermal steady-state is reached, as illustrated in the uncalibrated amplitude error after $\sim 10$ minutes of RB measurements in Fig.~\ref{fig:fig3}(c).
Interleaved calibrations reduce the amplitude drift error from $\sim 1.4 \times 10^{-7}$ to $9(7) \times 10^{-9}$.
The effectiveness of the amplitude calibration is limited by the amplitude resolution of the AWG, and by ``fast'' amplitude fluctuations.
The finite amplitude resolution of the AWG (15 bits) leads to a quantisation of the pulse amplitude which, on average, translates to a gate error of $\sim 1.5 \times 10^{-9}$.
We measure ``fast'' changes in amplitude between sequential shots of the experiment (i.e., on a $\sim$ 100 ms timescale) using the same procedure as for the amplitude calibration.
If the amplitude changes between sequential shots of the experiment, the net rotation induced by the pulse sequence will fluctuate, reducing the probability of recovering state $\ket{0}$ at the end of the sequence, as shown in Fig.~\ref{fig:fig3}(d).
We would observe the same decay if the amplitude were changing throughout the sequence of pulses.
To separate the changes in amplitude within a single sequence from shot-to-shot offsets, we repeat the same measurement but switch the phase of the $\pi/2$ pulses by $180^{\circ}$ halfway through the sequence.
We find that the error caused by drifts in amplitude during a sequence is $< 10^{-10}$, and is therefore negligible compared to shot-to-shot changes ($2.3(2) \times 10^{-8}$ error).
Further details can be found in~\citesec{SuppInfoLabel}{\ref{SI:amp_noise}}.

\begin{figure}[h]
\centering
\includegraphics[width=0.45\textwidth]{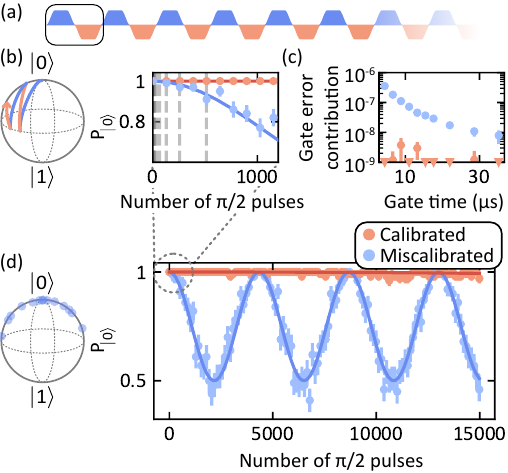}
\caption{
\textbf{Pulse frequency calibration.}
\textbf{(a)} The qubit frequency is subject to an ac Zeeman shift induced by the MW pulses.
This is measured by driving pairs of $\pi/2$ pulses where the second pulse has a phase $180^{\circ}$ relative to the first.
Ideally, the ion is returned to state $\ket{0}$ at the end of the sequence.
\textbf{(b)} The number of pulses is increased exponentially and, when the probability $P_{\ket{0}}$ of recovering the initial state is below a chosen threshold, the frequency miscalibration is computed and corrected.
\textbf{(c)} This procedure reduces the impact of ac Zeeman shifts from being a leading error contribution (blue), to being negligible (orange).
\textbf{(d)} If the frequency were changing between sequential shots of the experiment (as we observe for the pulse amplitude, Fig.~\ref{fig:fig3}(d)), undamped oscillations of $P_{\ket{0}}$ between 1 and 0.5 in the case of a miscalibrated frequency (blue), or a constant $P_{\ket{0}}=1$ for a well-calibrated frequency (orange), would not be achievable.
}
\label{fig:fig4}
\end{figure}

The second calibration required is that of the MW frequency.
The qubit is insensitive, to first-order, to dc Zeeman shifts, but remains sensitive to ac Zeeman shifts.
These may be ``ambient'' shifts, such as those caused by the radio-frequency trapping field, or ``MW-induced'' shifts caused by the gate drive.
Both types of ac Zeeman shift are calibrated before starting RB measurements.
The ``ambient'' shift is calibrated through a Ramsey measurement, with uncertainty $< 1$ Hz.
Miscalibration of this shift is not included in the error budget since the ``MW-induced'' shift calibration captures residual offsets in any ``ambient'' shifts.
The ac Zeeman shift induced by the MW pulses is calibrated by driving pairs of $\pi/2$ rotations around the $\pm$ X-axes on the Bloch sphere, where the second pulse has a phase shift of $180^{\circ}$ relative to the first, as shown in Fig.~\ref{fig:fig4}(a).
If the MW drive has the correct frequency, the second pulse returns the ion to its starting state.
If the MW drive is detuned with respect to the ac Zeeman-shifted qubit frequency, the pair of pulses will drive a net rotation about the Y-axis, see Fig.~\ref{fig:fig4}(b).
In practice, we append the sequence with a $\pi/2$ rotation around the Y-axis, such that a measurement of $P_{\ket{0}}=|\braket{\psi|0}|^{2}$ reveals a detuning offset as a deviation from $P_{\ket{0}}=1/2$.
With the MW amplitude set so as to produce a mean gate duration of $13\ \upmu$s, the ion experiences an ac Zeeman shift of $9(1)$ Hz which, if left uncorrected, would lead to a gate error of $6.0(7) \times 10^{-8}$, see Fig.~\ref{fig:fig4}(c).
Unlike the amplitude calibration, neither of these frequency calibrations is limited by ``fast'' changes in frequency or by the frequency resolution of the AWG ($6$ mHz).
To verify the former, we take the same approach as with the amplitude calibration scheme, and drive 15,000 pairs of $\pi/2$ pulses, observing none of the consequences expected from frequency fluctuations (see Fig.~\ref{fig:fig4}(d)).
If the frequency were changing on either a shot-to-shot or a pulse-to-pulse timescale, we would observe a decay in the probability of measuring $\ket{0}$ after many repetitions of $\pi/2$ pulse pairs.
Frequency noise is therefore found to be a negligible source of error; we choose to calibrate the MW frequency such that the impact on gate error is $< 1 \times 10^{-8}$.
There remains a varying detuning during the pulse ramp up/down and the inter-pulse delay -- since the ac Zeeman shift is dependent on the MW amplitude -- but the resulting error is $< 1 \times 10^{-9}$.

\begin{figure}[t]
\centering
\includegraphics[width=0.45\textwidth]{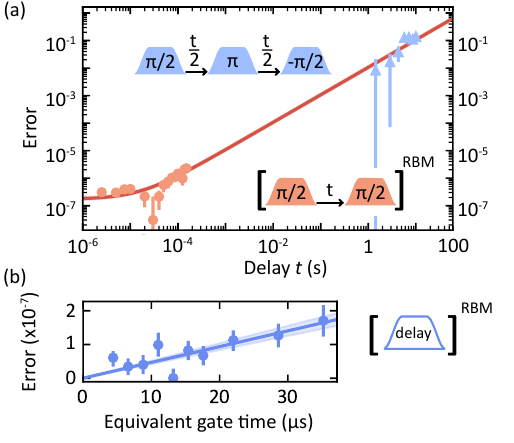}
\caption{
\textbf{Measurements of incoherent errors.}
\textbf{(a)} Decoherence errors are measured in an interleaved randomised memory benchmarking experiment \cite{omalley2015, sepiol2019} (orange circles), in which a delay $t$ is inserted between the $\pi/2$ pulses of a gate benchmarking sequence.
A linear fit to the data (red) yields a decoherence time constant $T_{2}^{**} = 69(7)$ s.
This is in good agreement with the error measured at much longer timescales by a spin-echo Ramsey measurement \cite{hahn1950, langer2005, biercuk2009, wang2021} (blue triangles).
\textbf{(b)} The combined impact of leakage, bit-flip and time-dependent measurement errors is estimated using randomised benchmarking (RB) measurements where we replace all $\pi/2$ pulses with a delay of the same duration.
Such measurements are collated here for delays corresponding to various gate times, and a linear fit is used to model this error.
}
\label{fig:fig5}
\end{figure}

Finally, we will discuss gate errors that are unrelated to MW pulse calibrations.
From Table~\ref{tab:error_budget}, we see that the dominant source of error is decoherence.
This is measured in two ways: by interleaved randomised memory benchmarking (IRMB) at short timescales \cite{omalley2015, sepiol2019} and by spin-echo Ramsey measurement at long timescales \cite{hahn1950, langer2005, biercuk2009, wang2021}, with both sets of results shown in Fig.~\ref{fig:fig5}(a).
During IRMB, the delay between $\pi/2$ pulses in a RB sequence is increased and the subsequent increase in gate error provides a measure of a decoherence time constant, equal to $69(7)$~s, that we denote by $T_{2}^{**}$.
The absence of the MW-induced ac Zeeman shift during the delay is accounted for by adjusting the phase of the subsequent $\pi/2$ pulses in the sequence.
Errors caused by leakage, bit-flip and measurement are subtracted to isolate the effect of decoherence, using the independent measurement shown in Fig.~\ref{fig:fig5}(b).
We determine the dominant source of decoherence to be phase noise in the MW drive chain~\cite{ball2016,cywinski2008} (further analysis can be found in~\citesec{SuppInfoLabel}{\ref{SI:decoherence}}).
Leakage, bit-flip and time-dependent measurement errors constitute the second-largest source of error.
These errors are measured using RB measurements, but with all MW pulses replaced with delays, see Fig.~\ref{fig:fig5}(b).
The error increases with delay due to: ``leakage'', population leaking out of the qubit subspace; ``bit-flips'', population transfer between the two qubit states; or ``time-dependent measurement errors'', an increase in the measurement error after a longer delay.
We observed significant improvements in the first two error sources by reducing undesired laser and MW radiation, see~\citesec{SuppInfoLabel}{\ref{SI:leakage}} for further details.
The remaining significant error source arises from the harmonic motion of the ion in its trapping potential, induced by motional heating during the gate sequence.
This modulates the pulse amplitude since the ion is positioned on a MW amplitude gradient engineered for driving two-qubit gates~\cite{weber2024}.
During a RB experiment, the in-plane radial motional mode of the ion is dark-resonance cooled to $2.6(8)$ quanta before state-preparation~\cite{allcock2016}, but heats up at a rate of $0.37(5)$ quanta/ms.
Through simulation \citesec{SuppInfoLabel}{\ref{SI:Errors}}, we estimate a gate error contribution due to this effect of $1.3(2) \times 10^{-8}$.
The coupling to the other motional modes is much lower due to a significantly smaller MW field gradient in the axial and out-of-plane directions, making the error contribution from these modes negligible.
In conclusion, we have achieved a new state-of-the-art single-qubit gate error of $1.5(4) \times 10^{-7}$ and presented a breakdown of all known sources of error, which agrees well with the data over a wide range of gate speeds.
Our fastest Clifford gate ($4.4~\upmu$s) has an error of $2.9(5)\times 10^{-7}$; this represents a $20\times$ speed-up relative to the previous lowest-error gate~\cite{harty2014}.
Compared with our previous work~\cite{harty2014}, we have reduced the error by nearly an order of magnitude by better control of microwave amplitude and detuning with automated calibration procedures, and by reduced excitation of spectator transitions using pulse shaping and larger Zeeman splittings.
Over the range of gate times studied, we determine the dominant contribution to the error to be decoherence, closely followed by leakage, bit-flip and time-dependent measurement errors.
The gate error could therefore be reduced further by improving the qubit coherence or by eliminating leakage out of the qubit states due to unwanted sources of radiation.
Alternatively, the gate error could be reduced using state leakage detection techniques~\cite{wu2022,kang2023,stricker2020}.
Calibrating the readout for different experimental sequence durations would allow us to separate the time-dependent measurement error from the gate error more unambiguously.
Amplitude and frequency noise of the MW pulses contribute an insignificant amount to the gate error, demonstrating the robustness of electronic control.
With hardware improvements, one could implement longer gate sequences, enabling experiments to probe the non-Markovian nature of the different sources of error~\cite{wallman2014,fogarty2015,mavadia2018,Figueroa_2021}.
Note that these types of errors could cause the RB measurements used here (and in many other similar experiments) to underestimate the gate error~\cite{Epstein_2014,ball2016RBM}.
For addressed single-qubit gates, multiple techniques exist to null the microwave field in non-addressed zones~\cite{warring2013,craik2017,loschnauer2024}, and even modest levels of cancellation can result in very low crosstalk through the use of composite pulses~\cite{merrill2014,loschnauer2024}.
Alternatively, this gate mechanism can drive different single qubit gates on $n$ ions within the same zone or potential well, using a sequence of $\sim 3n/2$ pulses~\cite{leu2023}.
For composite pulse sequences, or applications such as protecting memory qubits by dynamical decoupling, the relevant error would be the error per physical $\pi/2$ pulse, which is $0.7(2) \times 10^{-7}$.

\vspace{5mm}
\textbf{Acknowledgments}
The authors wish to thank A. Kwiatkowski, D.H. Slichter and R. Srinivas for insightful discussions; and J. Grover and H. Stemp for comments on the manuscript.
The trap was designed by T.P. Harty, with contributions from D.T.C. Allcock and D.M.L.
This work was supported by the U.S. Army Research Office (ref. W911NF-18-1-0340) and the U.K. EPSRC Quantum Computing and Simulation Hub (ref. EP/T001062/1).
M.C.S. acknowledges support from Balliol College, Oxford.
A.D.L. acknowledges support from Oxford Ionics Ltd.
K.M. acknowledges support from the Japan Science and Technology Agency (JST) ASPIRE Japan (ref. JPMJAP2319), and the JST Moonshot Research and Development program (ref. JPMJMS2063).

\vspace{5mm}
\textbf{Author Contributions}
M.C.S., A.D.L., M.F.G. and K.M. developed and maintained the experimental apparatus, and acquired the data.
M.C.S., A.D.L. and M.F.G. carried out the data analysis.
M.C.S. wrote the manuscript with contributions from all authors.
M.F.G. and D.M.L. supervised the work.

%% file: SI.tex
\section{Experimental setup}\label{SI:MW}

\begin{figure*}[hbtp]
    \centering
    \includegraphics[width=1\textwidth]{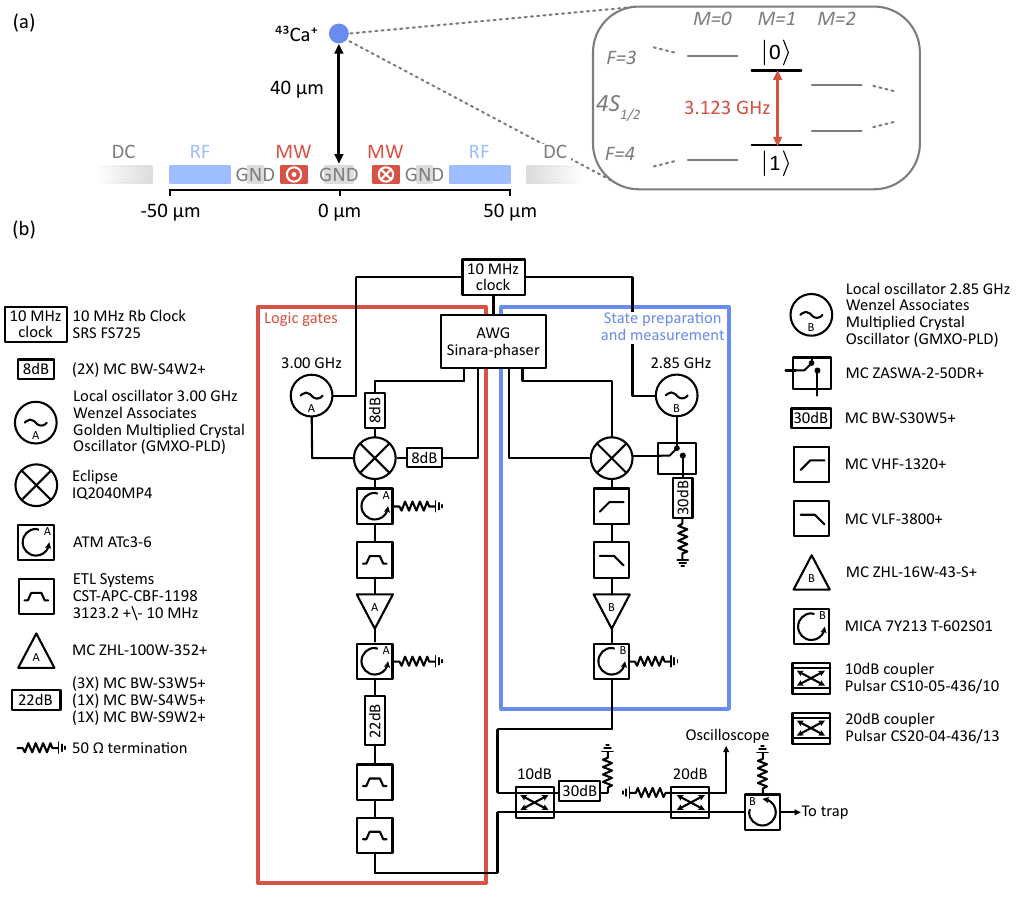}
    \caption{
    \textbf{System overview}
    \textbf{(a)} Surface trap electrode geometry and hyperfine structure (ion not to scale).
    Electrodes (thickness $4 \ \upmu$m) delivering radio-frequency and dc currents to generate a trapping potential are labeled RF and DC, microwave electrodes are labelled MW, and the electrical ground is labelled GND.
    Experiments are carried out using $^{43}$Ca$^{+}$ ions, in which quantum logic operations are driven using microwaves delivered on-chip at 3.123 GHz.
    \textbf{(b)} Microwave drive chain.
    The circuitry on the left (red box) produces pulses at the qubit frequency, used for driving quantum logic operations.
    The circuitry on the right (blue box) produces pulses in a wider range of frequencies, used for transferring the population around the $4^{2}S_{1/2}$ ground level manifold during state preparation and measurement.
    The two chains are combined before being delivered to the on-chip microwave resonator via a microwave vacuum feedthrough.
    }
    \label{fig:mw_chain}
\end{figure*}

Fig.~\ref{fig:mw_chain} shows an overview of the surface trap and microwave drive chain used in this experiment.
Further details on the surface trap chip and the qubit choice can be found in Ref.~\cite{weber2022}.

\section{Randomised benchmarking results}\label{SI:RBM}

\begin{figure}[h]
    \centering
    \includegraphics[width=1\textwidth]{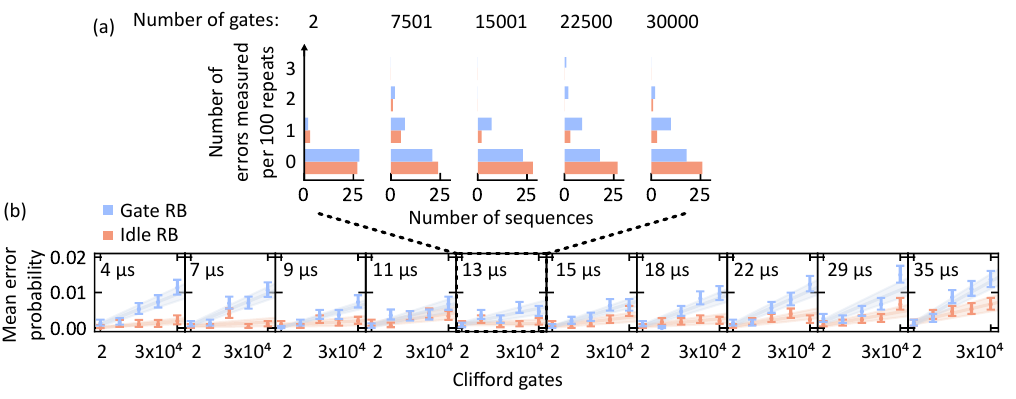}
    \caption{
    \textbf{(a)} Raw data underlying the RB measurement result shown in Fig.~\ref{fig:fig1}.
    For each of the 5 sequence lengths used in our measurements, we generate 30 pseudo-random gate sequences and repeat each one 100 times.
    The histogram shows the number of times an error was measured for all 30 gate sequences.
    Results are shown for gate RB measurements (blue) and also for ``idle RB'' measurements (orange), where all gate pulses are replaced by delays of the same duration.
    \textbf{(b)} Gate and idle RB measurement results for all gate durations.
    Gate RB errors resulting from these measurements are plotted in Fig.~\ref{fig:fig2}, and idle RB errors are shown in Fig.~\ref{fig:fig5}(b).
    }
    \label{fig:rbms_hist}
\end{figure}

In this section, we give further details on the randomised benchmarking (RB) results presented in Figs.~\ref{fig:fig1},\ref{fig:fig2}.
The ion is pseudo-randomly prepared in either state $\ket{0}$ or $\ket{1}$ before being subject to a pseudo-random series of Clifford gates.
We then pseudo-randomly shelve either $\ket{0}$ or $\ket{1}$ into the ``dark'' $3^{2}\text{D}_{5/2}$ manifold and measure the probability of the ion fluorescing.
The survival probability ($1-$ error probability) is defined as the probability of measuring the expected state at the end of the sequence.
For each of the 5 different sequence lengths used in our measurements, we generate 30 pseudo-random gate sequences (generated using~\cite{OITG}, list of sequences is available on request), and each is run 100 times.
We refer to each of these 100 instances as a single ``shot''.
The outcomes from each of these measurements are shown in Fig.~\ref{fig:rbms_hist}(a) for the optimal gate duration of 13 $\upmu$s.
The complete dataset was taken over the course of 6 days, with data taken each day for all 10 different gate durations (results for all gate durations measured are shown in Fig.~\ref{fig:fig2} and Fig.~\ref{fig:rbms_hist}(b)).
RB measurements were repeated but with all microwave pulses replaced by delays of the same duration to measure the leakage, bit-flip and time-dependent measurement error.

To extract gate error rates from the raw RB data, we use a maximum likelihood estimation (MLE) method.
This approach assumes that the survival probabilities follow a simple exponential decay as a function of the sequence length:
\begin{equation}
    p(l) = A\left(1-2\epsilon\right)^l+\frac12
    \label{eq:survival_model}
\end{equation}
where $l$ is the sequence length, $\epsilon$ is the depolarisation rate, and $A$ is a fit parameter which accounts for SPAM errors~\cite{knill2008}.
The error per Clifford gate is then given by $\epsilon$.
This model assumes that errors are Markovian which may not always be true.
A more rigorous analysis would involve fitting the data to a more general model, such as those presented in Refs.~\cite{wallman2014, fogarty2015}.
Since, in our measurements, the total error we accumulate at the maximum applied sequence length is only $4.7\times 10^{-3}$ (the maximum sequence length is limited by our hardware memory), we do not observe a significant deviation from the simple exponential model and so cannot investigate the presence of non-Markovian errors using our RB data.
Given the binary nature of the measurement outcomes (fluorescence or no fluorescence), we assume that the data follows a binomial distribution.
For each randomised sequence, we calculate the likelihood of obtaining the observed measurement outcome, and MLE is used to find the best-fitting parameters that maximise this likelihood.
To estimate the uncertainties in the fitted parameters, we employ parametric bootstrapping.
Using the fitted parameters from the MLE, we generate simulated datasets that follow the same binomial distribution.
The fitting procedure is repeated on these simulated datasets, yielding a distribution of the fit parameters.
The standard deviation of this distribution is then used to determine the uncertainties on our final reported values.

\section{Error budget construction}\label{SI:Errors}

Here we provide further details on the construction of the error budget presented in Table~\ref{tab:error_budget} and Fig.~\ref{fig:fig2}.
Each error is discussed in order of decreasing magnitude.
Analytical formulas for the error per $\pi/2$ pulse are provided where applicable, as a function of the pulse time $t_{\pi/2}$.
These can be converted to the error per Clifford gate by scaling them by the average number of $\pi/2$ pulses per Clifford gate, in our implementation $\simeq2.2$.

\subsection{Decoherence}
The data acquired through interleaved randomised memory benchmarking, shown in Fig.~\ref{fig:fig5}(a), is fitted following
\begin{equation}
    \epsilon_{\text{gate}} = \frac{2.2}{3}\frac{\left(t_{\pi/2} + t_{\text{delay}}\right)}{T_{2}^{**}} ,
    \label{eq:decoherence_fit}
\end{equation}
where $t_{\pi/2}=10\upmu$s, and $\epsilon_{\text{gate}}$ is the gate error after subtracting the errors caused by leakage, bit-flip and measurement, to determine the decoherence timescale $T_{2}^{**}$.
The average error per $\pi/2$ pulse is then well approximated by:
\begin{equation}
    \epsilon_{T_{2}} \simeq \frac{1}{3}\frac{t_{\pi/2}}{T_{2}^{**}} .
\end{equation}
where the factor $1/3$ arises due to averaging the qubit state over the Bloch sphere.

\subsection{Leakage, bit-flip and time-dependent measurement}
Errors were measured through randomised benchmarking experiments, where all microwave pulses were replaced by a delay of the same duration, see Sec.~\ref{SI:RBM} for more information.
Further measurements can be found in Sec.~\ref{SI:leakage}.

\subsection{Harmonic motion}
Our system is designed to maximise the gradient of the Rabi frequency in the in-plane radial direction for fast entangling operations~\cite{weber2022,weber2024}.
Consequently, this means that heating of the in-plane radial mode will cause the Rabi frequency to fluctuate at the motional frequency $\omega_m=2\pi\times5.6$ MHz, inducing single-qubit gate errors.
A derivation of the relevant Hamiltonian, as well as a detailed characterisation of experimental parameters for our system, can be found in the supplementary information of Ref.~\cite{smith2024}.
The key metric is the effective Lamb-Dicke parameter $\eta=9.3\times 10^{-4}$, quantifying the ion-motion coupling strength relative to the Rabi frequency.
Using the Hamiltonian of Ref.~\cite{smith2024}, we simulate the average error (as defined in Ref.~\cite{bowdrey2002}) for a $\pi/2$ pulse, for different values of $\bar n$, where $\bar n$ is the average thermal occupation of the motional mode.
We restrict these simulations to the regime $\bar n < 2$, such that only $n<20$ Fock states of the motional mode need to be used.
We observe an oscillating error as the pulse duration is varied, where no error is found for pulse times matching a multiple of the motional oscillation period.
We then empirically construct an approximate analytical expression for the envelope $\epsilon_m$ of these error oscillations:
\begin{equation}
\begin{split}
    \delta\Omega &= \frac{3\pi}{4}\tfrac{\eta\sqrt{\bar n+\frac12}}{t_{\pi/2}}\ ,\\
    \epsilon_m &\simeq \left(\frac{\delta\Omega}{\omega_m}\right)^2\ .
\end{split}
\end{equation}
We find that the change in $\bar n$ during a pulse has a negligible effect and that our pulse shaping is too short to mitigate motional errors.
Since we expect $\bar n$ to increase linearly over a gate sequence (our heating rate is measured to be $\dot{\bar{n}} =370(50)$ quanta per second), we quote the average error over the sequence for the longest sequence used (30,000 gates).

This formula was determined for $\bar n<2$, much smaller than the highest mode occupations of $\bar n\sim 300$ reached in the longest RB sequences.
We therefore carried out a quasi-classical simulation to validate the expression for $\epsilon_m$ at higher $\bar n$.
In this simulation, we consider that every pulse is subject to amplitude modulation with a random phase, a frequency given by the in-plane radial mode frequency, and a modulation amplitude of $2\eta\sqrt{\bar n}$ relative to nominal pulse amplitude.
We also find excellent agreement between the resulting error and the empirically determined formula for $\epsilon_m$.

\subsection{Amplitude noise}
As discussed in detail in Sec.~\ref{SI:amp_noise} and Fig.~\ref{fig:fig3}, there is a significant shot-to-shot variation $\delta\Omega$ in the Rabi frequency $\Omega$.
For such coherent pulse errors, we find that the pulse error defined in Ref.~\cite{bowdrey2002} incorrectly estimates the error measured with RB.
We therefore empirically approximate the error of a $\pi/2$ pulse, based on RB measurement simulations, to be:
\begin{equation}
\epsilon_{\delta\Omega} = \frac{1}{2}\left(\frac{\delta\Omega}{\Omega}\right)^2\ .
\label{eq:amplitude_error}
\end{equation}
The average error due to amplitude noise is then determined using this expression in conjuction with the measured amplitude distribution determined following Sec.~\ref{SI:amp_noise}.

\subsection{Amplitude drift}
Amplitude calibrations are interleaved with our RB measurements.
This serves a dual purpose: to maintain an optimal pulse area and thus reduce the gate error, but also to provide an accurate characterisation of drifts in amplitude.
We assume a linear drift in amplitude between two setpoints of the calibration experiment, and use Eq.~\ref{eq:amplitude_error} to construct an estimate of this error for every RB measurement run.
These errors are combined to estimate the average error contribution.
\subsection{ac Zeeman shift}
The ac Zeeman shift is measured, and subsequently corrected for, through the procedure illustrated in Fig.~\ref{fig:fig4}.
The process is repeated to verify that the residual Zeeman shift will not lead to an error exceeding a predetermined threshold.
The error $\epsilon_Z$ caused by this residual Zeeman shift $\Delta_Z$ is the number quoted in our error budget.
As with amplitude errors, the coherent nature of this error requires us to empirically establish an analytical expression of the error, based on simulations of a RB measurement.
We find the resulting error per $\pi/2$ pulse is well approximated by:
\begin{equation}
    \epsilon_Z \simeq 2\pi(t_{\pi/2}\Delta_Z)^2\ .
\end{equation}

\subsection{Amplitude resolution}
The AWG (Sinara Phaser) has 15 bits of amplitude resolution, which leads to a quantisation of the pulse amplitude.
In practice, the amplitude is programmed through an amplitude scale factor $\in[0,1]$, rounded to the nearest multiple of $1/2^{15}$.
Writing the optimal amplitude as $a$, the closest optimal amplitude setting for the AWG lies in the range $\left[a-0.5/2^{15},a+0.5/2^{15}\right]$, leading to a relative Rabi frequency offset in the range $\left[-1/(2^{16}a),+1/(2^{16}a)\right]$.
Assuming we randomly sample Rabi frequency offsets in this range, using Eq.~\ref{eq:amplitude_error} the quantisation error for a $\pi/2$ pulse is given by:
\begin{equation}
    \epsilon_{\text{AWG}} = \frac16\left(\frac{1}{2^{16}a}\right)^2\ .
\end{equation}

\subsection{Excitation of spectator states}
The qubit states are dressed by the four spectator states $\ket{F, M}_{F\in[3,4],M\in[0,2]}$ (Fig.~\ref{fig:mw_chain}(a)) during a gate.
The spectator transitions are each detuned, on average, by $\Delta_S\approx104$ MHz from the qubit transition.
The Rabi frequencies of these transitions $\Omega_S$ are, on average, 1.4$\times$ larger than the qubit Rabi frequency, taking into account both the difference in the magnetic dipole moment (0.7$\times$) relative to that of the qubit transition, and the amplitude of the $\sigma$-polarised magnetic field (2$\times$), relative to the $\pi$-polarised field driving the qubit.
The presence of these states can affect the gate error in two ways.
Firstly, population can remain within these states at the end of a pulse, causing leakage out of the qubit subspace.
The impact of this is determined by the pulse ramp time (40 ns in our case) relative to the effective Rabi period for these spectator transitions ($\approx 1/\Delta_{s} \approx 10$ ns).
Secondly, the portion of the population residing in these spectator states is subject to increased decoherence, resulting from the first-order magnetic field dependence of the spectator transitions.
We measure a decoherence time $T_{2,S}^{**}\sim40$ ms for these transitions using interleaved randomised memory benchmarking, which is three orders of magnitude less than the qubit coherence time.
However, this reduced coherence affects only a small proportion of the population residing in each state $\sim \Omega_S^2/(\Delta_S^2+\Omega_S^2)\lesssim 3\times 10^{-6}$.
Even at the shortest pulse times used in this work (2 $\upmu$s), both these effects only amount to an error of $\simeq1.5\times10^{-9}$ when simulating RB sequences in the presence of all spectator states.
At the optimal gate duration of $13$ $\upmu$s, the error is well below $10^{-9}$ per gate.
\subsection{Inter-pulse Zeeman shift ramping}
We simulate RB sequences, in which the pulses are ramped up and down over 40 ns, and include a 40 ns inter-pulse delay.
The varying ac Zeeman shift, characterised following Fig.~\ref{fig:fig4}, results in an error upper-bounded to $10^{-9}$.

\subsection{Non-rotating-wave dynamics}
A complete model of the microwave driving, on resonance and in the interaction picture, is given by
    \begin{equation}
        H = \frac{\Omega\left(t\right)}{2} \left(e^{-i\varphi}\hat{\sigma}_{-}+e^{-2i\omega_q t}e^{-i\varphi}\hat{\sigma}_{-} + \text{h.c.}\right)\ ,
    \end{equation}
where $\Omega\left(t\right)$ is the Rabi frequency, $\varphi$ the microwave phase, and $\omega_q$ the qubit frequency.
Whilst we neglect the term oscillating with frequency $2\omega_q$ in other simulations by invoking the rotating wave approximation (RWA), here we have simulated its impact on the gate error, and find it to be lower than $10^{-10}$.

\section{Fluctuations in pulse amplitude}\label{SI:amp_noise}

In this section, we discuss the results presented in Fig.~\ref{fig:fig3} and provide further measurements supporting the claim that the amplitude is changing between sequential shots of the experiment ($\sim$ 100 ms timescale).
\subsection{Reproducible amplitude offset}
\label{SI:amp_noise:amp_calibration}
The Hamiltonian describing the interaction between the qubit and microwaves can be written in the interaction picture as
\begin{equation}
    H = \frac{\Omega\left(t\right)}{2} \hat{\sigma}_{x}\ ,
\end{equation}
where $\Omega\left(t\right)$ is the Rabi frequency.
We have assumed that the microwave phase is constant and that there is no detuning between the microwave and qubit frequency.
Starting in state $\ket{0}$, the state of the qubit after some time $t$ is then
\begin{equation}
    \ket{\psi\left( t \right)} = \exp\left(-\frac{\text{i}}{2} \int_{0}^{t} d t' \ \Omega\left(t'\right) \hat{\sigma}_{x}\right) \ket{0}\ .
\end{equation}
In the ideal case of a fixed Rabi frequency $\Omega\left(t\right)=\Omega_{q}$ this describes Rabi oscillations between the two qubit states $\ket{0}$ and $\ket{1}$.
By choosing to drive $4N$ $\pi/2$ pulses, i.e. a total drive time of
\begin{equation} \label{eq:four_pihalf_pulses}
    t=2\pi N/\Omega_{q}\ ,
\end{equation}
the final state of the qubit is
\begin{equation} \label{eq:evolution_static_offset}
    \ket{\psi} = \exp\left(-\text{i}  \pi N \hat{\sigma}_{x}\right) \ket{0}= \left(-1\right)^{N}\ket{0}\ ,
\end{equation}
meaning the qubit undergoes $N$ perfect $2\pi$ rotations.
Now let us consider a miscalibration $\Omega_{0}$ in the pulse amplitude such that there is a constant offset in Rabi frequency $\Omega\left(t\right)=\Omega_{q} + \Omega_{0}$.
The state of the qubit at the end of the sequence of $\pi/2$ pulses will be
\begin{equation}\label{eq:amp_oscill}
    \ket{\psi} = \left(-1\right)^{N} \exp\left(-\text{i}  \pi N \frac{\Omega_{0}}{\Omega_{q}} \hat{\sigma}_{x}\right) \ket{0} .
\end{equation}
Measuring the probability $P_{\ket{0}}$ of finding state $\ket{0}$ at the end of the sequence reveals the offset $\Omega_0$ following
\begin{equation}
    P_{\ket{0}} = |\braket{0|\psi}|^2= \cos^{2} \left(\pi N \frac{\Omega_{0}}{\Omega_{q}}\right)\ .
\end{equation}

\subsection{Changes in amplitude between measurements}

We now consider the case in which the offset is randomly sampled from a Gaussian distribution every shot of the experiment:
\begin{equation}
    \text{Prob}\left(\Omega_{0}\right) = \frac{1}{\sigma_{0}\sqrt{2\pi}} \exp\left(-\frac{\left(\Omega_{0}-\mu_{0}\right)^{2}}{2 \sigma_{0}^{2}}\right)\ ,
\end{equation}
where $\mu_{0}$ is the mean and $\sigma_{0}$ is the standard deviation of the amplitude offset $\Omega_{0}$.
To determine the probability $P_{\ket{0}}$ of finding state $\ket{0}$ at the end of the sequence, we integrate over the distribution of amplitude offsets, yielding
\begin{equation}
\begin{split}
    P_{\ket{0}} &= \int_{-\infty}^{\infty} d \Omega_{0}  \frac{1}{\sigma_{0}\sqrt{2\pi}} \exp\left(-\frac{\left(\Omega_{0}-\mu_{0}\right)^{2}}{2 \sigma_{0}^{2}}\right) \cos^{2} \left(\pi N \frac{\Omega_{0}}{\Omega_{q}}\right)  \\
    &=  \frac{1}{2}\cos \left(2\pi N \frac{\mu_{0}}{\Omega_{q}}\right) \exp\left(-\frac{1}{2}\left(2\pi N \frac{\sigma_{0}}{\Omega_{q}}\right)^{2}\right) + \frac{1}{2}\ .
\end{split}
\end{equation}
A mean offset thus produces sinusoidal oscillations in $P_{\ket{0}}$, whilst shot-to-shot variations in amplitude cause a decay of $P_{\ket{0}}$.
Both effects are visible in the data presented in Fig.~\ref{fig:fig3}(d).

\subsection{Changes in amplitude during a single shot of a sequence}

The microwave amplitude may also be changing during a single shot of the sequence of $\pi/2$ pulses.
To describe this effect, we must consider a time-dependent amplitude offset
\begin{equation}\label{eq:amp_polynomial}
    \begin{split}
        H &= \left(\Omega_{q}+\Omega_{0}+\Omega_{1}t +\Omega_{2}t^2+...\right)\frac{\hat{\sigma}_{x}}{2}\\
        &= \left(\Omega_{q} + \sum_{k=0}^{\infty} \Omega_{k} t^{k}\right)\frac{\hat{\sigma}_{x}}{2}\ .
    \end{split}
\end{equation}
Following the same approach as above in Sec.~\ref{SI:amp_noise:amp_calibration}, the evolution of the qubit state after $4N$ $\pi/2$ pulses is described by
\begin{equation} \label{eq:state_amp_polynomial}
    \ket{\psi} =  \left(-1\right)^{N} \exp\left(\sum_{k=0}^{\infty}-\frac{\text{i}}{2} \frac{\Omega_{k}}{\left(k+1\right)} \left(\frac{2\pi N}{\Omega_{q}}\right)^{k+1} \hat{\sigma}_{x}\right) \ket{0} .
\end{equation}

\subsection{Characterisation of amplitude changes using Walsh sequences}

\begin{figure}[h]
    \centering
    \includegraphics[width=0.9\textwidth]{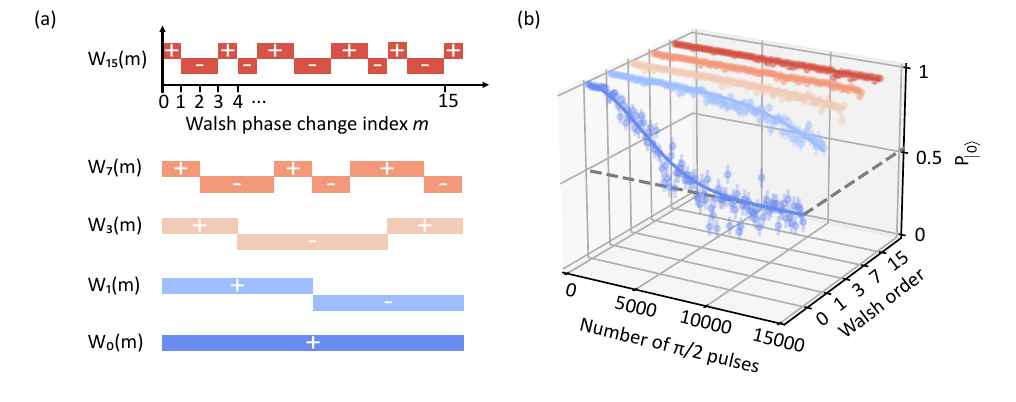}
    \caption{
    \textbf{Characterisation of amplitude variations during pulse sequences.}
    \textbf{(a)} Changes in the pulse amplitude are measured by driving multiples of four $\pi/2$ pulses such that deviations from the ideal pulse amplitude are detected as a net rotation of the qubit on the Bloch sphere.
    To diagnose the timescale of these changes in amplitude, we repeat this measurement but switch the phase of the $\pi/2$ pulses by $\pi$ radians at some point(s) during the sequence of pulses following a Walsh pattern.
    Here we show the different Walsh patterns implemented, defined by ``Walsh orders'' 0, 1, 3, 7 and 15.
    Switching the pulse phase following Walsh order 1 will cancel the effect of constant offsets in pulse amplitude.
    Walsh order 3 cancels linear changes, order 7 quadratic changes, and order 15 cubic changes.
    \textbf{(b)} Probability $P_{\ket{0}}$ of recovering the initial state after a train of $\pi/2$ pulses with phases switched following the Walsh sequences.
    The data is fitted following Eq.~\ref{eq:walsh_amp} to characterise the time-dependence of amplitude variations.
    The grey dashed line indicates where $P_{\ket{0}}=0.5$.
    }
    \label{fig:walsh_wiggles}
    \end{figure}

To determine the coefficients $\Omega_k$ describing the evolution of the microwave amplitude during a sequence, we repeat the measurements shown in Fig.~\ref{fig:fig3} but switch the phase of the $\pi/2$ pulses by $\pi$ rad at some points in time (indexed by $m$) following a Walsh sequence~\cite{beauchamp1984, hayes2012}.
The Walsh sequence is labelled by a Walsh ``order'' which can take values 0, 1, 3, 7, 15 or more generally $2^M-1$ with $M\in\mathbb{N}$.
For a given sequence order, the pulse phase is updated to $W_{2^M-1}(m)$ when the number of $4\times$ $\pi/2$ pulses reaches
\begin{equation}
    n_{m} = m \frac{N}{2^{M}}\ .
\end{equation}
The different Walsh sequences are defined in Fig.~\ref{fig:walsh_wiggles}(a) and Ref.~\cite{beauchamp1984}.
We will first consider the simplest sequence, Walsh order 1, where the phase of the $\pi/2$ pulses is changed by $\pi$ rad halfway through the sequence, and apply this to the case of a constant offset in the microwave amplitude.
By making use of the state evolution described by Eq.~\ref{eq:amp_oscill}, we find the qubit state at the end of the $N\times$ four $\pi/2$ pulses to be
\begin{equation}
\begin{split}
    \ket{\psi} &= \left(-1\right)^{-\left(n_{2}-n_{1}\right)} \exp\left(+\text{i}  \pi \frac{\Omega_{0}}{\Omega_{q}} \left(n_{2} - n_{1} \right) \hat{\sigma}_{x}\right) \left(-1\right)^{\left(n_{1}-n_{0}\right)} \exp\left(-\text{i}  \pi \frac{\Omega_{0}}{\Omega_{q}} \left(n_{1} - n_{0} \right) \hat{\sigma}_{x}\right) \ket{0} \\
    &= \exp\left(-\text{i}  \pi \frac{\Omega_{0}}{\Omega_{q}} \left(-n_{2} + 2 n_{1} - n_{0} \right) \hat{\sigma}_{x}\right) \ket{0} \\
    &= \ket{0}
\end{split}
\end{equation}
which shows that applying a Walsh order 1 sequence cancels the net rotation induced by constant offsets in amplitude.
When applying an arbitrary Walsh sequence to the arbitrary time-varying amplitude defined in Eq.~\ref{eq:amp_polynomial}, we obtain
\begin{equation}
    \begin{split}
    \ket{\psi} &=  \exp\left(-\frac{\text{i}}{2} \Omega_{q} A_{0} \hat{\sigma}_{x} + \sum_{k=0}^{\infty}-\frac{\text{i}}{2} \Omega_{k} A_{k} \hat{\sigma}_{x}\right) \ket{0}\ \text{and}\ P_{\ket{0}} = \cos^{2}\left(\sum_{k=0}^{\infty}\frac{\Omega_{k} A_{k}}{2}\right)\ ,\\
    A_{k} &= \frac{1}{\left(k+1\right)} \left(\frac{2\pi N}{\Omega_{q} 2^{M}}\right)^{k+1} \quad \sum_{m=0}^{2^{M}-1} W(m) \left(\left(m+1\right)^{k+1} - m^{k+1}\right)\ ,
    \end{split}
    \label{eq:walsh_evolution}
\end{equation}
where $W(m)=W_{2^M-1}(m)$ is the Walsh function of order $2^M-1$.
Using the definition of $W(m)$ and Eq.~(\ref{eq:walsh_evolution}), one can derive how a Walsh-1 sequence cancels the effect of constant offsets in amplitude ($A_0=0$), Walsh-3 cancels linear changes ($A_0=A_1=0$), Walsh-7 cancels quadratic changes ($A_0=A_1=A_2=0$) and so on.
Furthermore, if the parameters $\Omega_{k}$ vary from shot to shot according to a Gaussian distribution:
\begin{equation}
    \text{Prob}\left(\Omega_{k}\right) = \frac{1}{\sigma_{k}\sqrt{2\pi}} \exp\left(-\frac{\left(\Omega_{k}-\mu_{k}\right)^{2}}{2 \sigma_{k}^{2}}\right)\ ,
\end{equation}
integrating over $\Omega_{k}$ leads to the following probability of recovering the initial state
\begin{equation}\label{eq:walsh_amp}
    P_{\ket{0}} =  \frac{1}{2} + \frac{1}{2}\prod_{k=0}^{\infty} \cos\left(A_{k} \mu_{k} \right) \exp\left(- \frac{1}{2}\left(A_{k} \sigma_{k}\right)^{2}\right)\ .
\end{equation}
To determine the timescale of amplitude fluctuations, we fit Eq.~\ref{eq:walsh_amp} to experimental data with Walsh sequences up to order 15 (see Fig.~\ref{fig:walsh_wiggles}).
Since Walsh sequences of orders $\geq 1$ almost completely cancel any net rotation induced after 15,000 $\pi/2$ pulses, we are unable to distinguish constant changes $\mu_{k}$ from shot-to-shot variations $\sigma_{k}$ in the parameters $\Omega_{k}$.
We therefore choose to fix $\sigma_{k}=0$ for $k\geq1$ and fit the data with only one free parameter $\mu_{k}$ for each Walsh order $\geq 1$.
We obtain errors of the same order of magnitude if we choose to fix $\mu_{k}=0$ and fit $\sigma_{k}$ instead.
In each successive fit, we ignore higher order contributions -- e.g. for the fit to Walsh-3 data we only consider oscillations due to $\mu_{2}$ and ignore all other $\mu_{k}$ -- which provides an upper-bound of the size of the amplitude variations.
We find that the dominant source of amplitude fluctuations is encompassed by $\sigma_{0}$: shot-to-shot variations in amplitude which remain constant during the pulse sequence.
The next largest source of error, linear variations in amplitude $\mu_{1}$, induce a negligible gate error of $< 10^{-10}$ per gate.
We note that similar methods have previously been developed to perform noise spectroscopy of quantum systems~\cite{ball2015,soare2014}.

\section{Sources of Decoherence}\label{SI:decoherence}

In this section, we discuss the different sources of decoherence present in our experimental setup and compare them to the measurements in Fig.~\ref{fig:fig5}(a).

\subsection{Phase noise in the microwave drive chain}

\begin{figure*}[hbtp]
    \centering
    \includegraphics[width=1\textwidth]{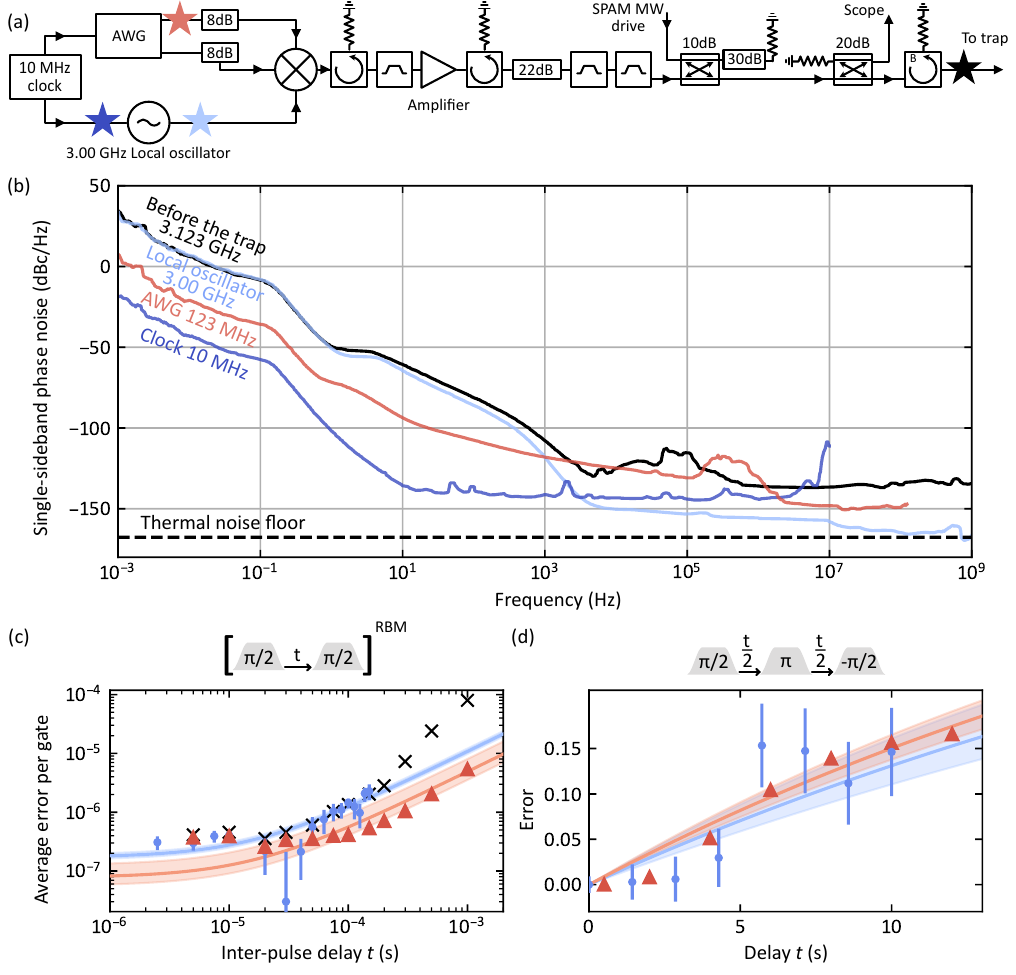}
    \caption{
    \textbf{Phase noise in the microwave drive chain.}
    \textbf{(a)} A simplified schematic of the microwave drive chain used to generate pulses at the qubit frequency.
    A more detailed schematic can be found in Fig.~\ref{fig:mw_chain}(b).
    \textbf{(b)} The single-sideband (SSB) phase noise measured at different points along the microwave drive chain, indicated by stars in (a).
    At low frequencies ($<1$ Hz) the phase noise is dominated by the local oscillator and at high frequencies ($>10$ kHz) the phase noise of the amplifier determines the noise floor, which is well above the level set by thermal noise (dashed line).
    Filter-transfer functions describing \textbf{(c)} interleaved randomised memory benchmarking (IRMB) and \textbf{(d)} spin-echo Ramsey measurements, were used in combination with the SSB phase noise measurements to determine the expected error during these measurements (red triangles).
    Comparing to experimental results (blue circles), shown in Fig.~\ref{fig:fig5}(a), indicates that phase noise plays a significant role in our observed decoherence.
    The phase noise measurements lead to excellent agreement in the low inter-pulse delay regime of IRMB, closest to the RB measurement.
    Other effects could lead to inflated errors at longer delays such as a slight miscalibration of the bare qubit frequency by $\sim 2.5$ Hz; accounting for this detuning leads to better agreement with the IRMB measurement at all delays (black crosses).
    }
    \label{fig:phase_noise}
\end{figure*}

Previous studies have shown that phase noise is the dominant source of decoherence in many trapped-ion experiments~\cite{sepiol2019, ball2016, wang2021, day2022}.
To determine the phase noise in our microwave drive chain, we measured the single-sideband (SSB) phase noise at different points along the chain using a phase noise analyser (Rohde $\&$ Schwarz FSWP), as shown in Fig.~\ref{fig:phase_noise}.
At low frequencies ($<1$ Hz) the phase noise is dominated by noise from the local oscillator.
We use a multiplied crystal oscillator, which converts a $10$ MHz Rb clock signal to $3.00$ GHz.
For an ideal oscillator, this process would amplify the SSB phase noise in the clock by $20 \log_{10}\left(\omega_{\text{LO}}/\omega_{\text{clock}}\right) = 50~$dB~\cite{ramon2013}, which is consistent with the measured local oscillator phase noise.
The low-frequency noise could therefore be reduced by improving clock phase noise, and/or reducing the noise amplification by using type of oscillator, for example a cryogenic sapphire oscillator~\cite{tan2023, hartnett2010}.
At high frequencies ($>10$ kHz) the phase noise of the amplifier dominates the noise.
We determine the white noise floor set by the amplifier to be well above the level set by room-temperature thermal noise (electronic noise generated by the thermal movement of charge carriers), indicated by the dashed line in Fig.~\ref{fig:phase_noise}(b).
The thermal noise floor of the SSB phase noise (in units of dBc/Hz) is calculated using
\begin{equation}
    \mathcal{L}_{\text{min}} = 30 + 10 \text{log} \left( k_{\text{B}} T \right) - P_{\text{carrier}}
\end{equation}
where $k_{\text{B}}$ is Boltzmann's constant, $T$ is the temperature in K, and $P_{\text{carrier}}$ is the power of the carrier frequency in units of dBm~\cite{ball2016, miller2012}.
The influence of phase noise on coherence measurements can be determined using filter transfer functions~\cite{ball2016,cywinski2008,miller2012}.
The measured SSB phase noise must first be transformed into a (unilateral) phase power spectral density
\begin{equation}
    \mathcal{S}_{\phi}\left(\omega\right) = 2 \times 10^{\mathcal{L}\left(\omega\right)/10}
\end{equation}
where $\mathcal{L}\left(\omega\right)$ is the SSB phase noise~\cite{lance1984, miller2012}.
The fidelity of a sequence of qubit operations as a function of time is then determined by
\begin{equation}
    \mathcal{F}\left(\tau\right) = \frac{1}{2}\left(1 + e^{-\chi\left(\tau\right)}\right)
\end{equation}
where $\chi\left(\tau\right)$ is the overlap integral of the phase power spectral density with a filter transfer function describing the sequence of qubit operations
\begin{equation}
    \chi\left(\tau\right) = \frac{1}{\pi} \int_{0}^{\infty} d \omega \ \mathcal{S}_{\phi}\left(\omega\right) G\left(\omega, \tau\right) .
    \label{eq:filter_function_overlap}
\end{equation}
We used an existing Python package to generate filter transfer functions~\cite{hangleiter2021} describing the decoherence measurements presented in Fig.~\ref{fig:fig5}(a).
A description of how to derive filter transfer functions for sequences of unitary qubit operations can be found in Refs.~\cite{hangleiter2021,green2013}.
To simulate IRMB measurements, we generated filter transfer functions to model different random sequences of up to $10,000$ Clifford gates.
We then computed the expected error for each sequence using Eq.~\ref{eq:filter_function_overlap}, and fitted the results using Eq.~\ref{eq:survival_model} to determine the IRMB error shown in Fig.~\ref{fig:phase_noise}(c).
For spin-echo Ramsey measurements, we generated filter transfer functions for a range of delays and computed the expected error using Eq.~\ref{eq:filter_function_overlap}, shown in Fig.~\ref{fig:phase_noise}(d).
Filter transfer functions for spin-echo experiments can also be written analytically~\cite{ball2016,cywinski2008}, which we used to verify our model.
Comparing the resulting ``simulated'' data indicates that phase noise in the microwave drive chain is a significant source of decoherence in our system.

\subsection{Magnetic field fluctuations}

Another possible source of decoherence is fluctuations in the static magnetic field.
To assess their impact, we first measured the frequency of the qubit transition as a function of magnetic field, confirming we are operating at a field where the qubit has a minimal first-order sensitivity to ``fast'' magnetic field fluctuations.
This also allows us to determine the change in sensitivity in the event of ``slow'' drift in the magnetic field mean value.
We then measured the decoherence time $T_{2}^{*}$ of a first-order magnetic field sensitive ``stretch'' transition ($\ket{F = 4, M = 4}\leftrightarrow\ket{F = 3, M = 3}$), using a spin-echo Ramsey measurement, to be $1.9(2)$ ms, revealing ``fast'' fluctuations on the order of $\sim 10$ nT.
Between each of our RB measurements, we also calibrate the static magnetic field by measuring the change in the frequency of the stretch transition, revealing ``slow'' changes in the magnetic field of $\sim 80$ nT.
These ``fast'' and ``slow'' changes in the magnetic field, combined with measurements of the qubit sensitivity, provide an estimate of the decoherence time as $\sim 3700$~s, making these effects negligible relative to the measured decoherence and the decoherence induced by microwave phase noise.

\section{Leakage, bit-flip and time-dependent measurement errors}\label{SI:leakage}
\begin{figure*}[h]
    \centering
    \includegraphics[width=0.9\textwidth]{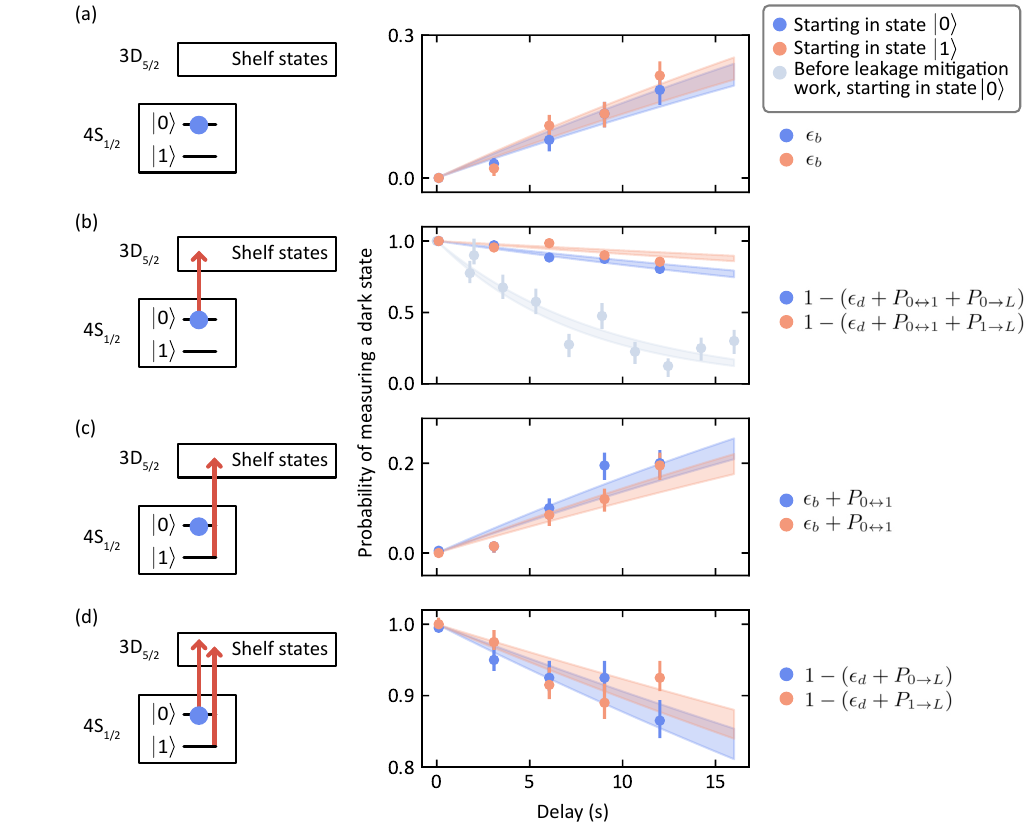}
    \caption{
    \textbf{Diagnostic measurements of measurement, leakage and bit-flip errors.}
    In all measurements presented, we prepare the ion in a qubit state ($\ket{0}$ or $\ket{1}$), wait for a varying delay, and shelve a chosen set of states which will be rendered dark in state-dependent fluorescence readout.
    The shelving scheme varies across measurements as follows: in \textbf{(a)} we measure the probability of the ion fluorescing after a delay, without shelving any states; in \textbf{(b)} we shelve the initially prepared state; in \textbf{(c)} we shelve the qubit state that was not prepared; and in \textbf{(d)} we shelve both qubit states.
    For each type of measurement, we measure after preparing $\ket{0}$ (blue) or after preparing $\ket{1}$ (red).
    The measurement, leakage and bit-flip error contributions for each measurement are shown in the formulas on the right.
    For (b), we also show the same measurement, starting in $\ket{0}$, before certain improvements were made to reduce leakage errors (grey).
    These measurements were combined to estimate the probability of a bit-flip error occurring, see Fig.~\ref{fig:bitlip}.
    }
    \label{fig:leakage}

\end{figure*}

The leakage, bit-flip and time-dependent measurement errors presented in Fig.~\ref{fig:fig5} and Table~\ref{tab:error_budget} were measured by repeating RB measurements without injecting microwaves, as described in Sec.~\ref{SI:RBM}.
The measurement effectively consists of state preparation, a delay of up to one second, and then readout.
Here we describe further measurements of this nature, notably extending the delays above ten seconds.
We thereby illustrate the difficulty in separating qubit leakage and qubit measurement errors, show their non-linear evolution with time and illustrate how hardware improvements suppressed leakage and bit-flip errors.
We separate the errors that are present without microwave driving into three categories:
\begin{enumerate}
    \item Measurement errors which increase with sequence duration, for example due to the Doppler broadening of the ion's response to the readout lasers as the ion's harmonic motion heats up. We distinguish the probability of dark measurement errors $\epsilon_d$ (e.g. shelving failures) from bright measurement errors $\epsilon_b$ (e.g. fluorescence detection failure).
    \item Leakage out of the qubit subspace, which we quantify with probabilities $P_{0 \rightarrow L}$ and $P_{1 \rightarrow L}$ for leakage out of states $\ket{0}$ and $\ket{1}$ respectively. These errors can for example be caused by undesired 397 nm or 393 nm photon scattering.
    \item Bit flip errors, occuring with probability $P_{0 \leftrightarrow 1}$. These can for example be driven by microwave noise at the qubit frequency. The assumption that $P_{0 \leftrightarrow 1} = P_{1 \rightarrow 0} = P_{0 \rightarrow 1}$ is reflected in the symmetry of the results in Fig.~\ref{fig:leakage}(c).
\end{enumerate}
When conducting RB measurements, we randomise the final state ($\ket{0}$ or $\ket{1}$), as well as the readout method (shelving the expected state or the other state).
Therefore, the above mentioned errors contribute to the RB error as follows:
\begin{equation}
    \epsilon_{\text{RB}} = \underbrace{\frac{1}{2}\epsilon_b}_{\text{Bright measurement error}}+\underbrace{\frac{1}{2}\left(\epsilon_d+\frac12 P_{0 \rightarrow L}+\frac12 P_{1 \rightarrow L}\right)}_{\text{Leakage and dark measurement error}}+\underbrace{P_{0 \leftrightarrow 1}}_{\text{Bit-flip error}}
    \label{eq:epsilon_RB}
\end{equation}
To measure these individual error sources, we prepare the qubit in state $\ket{0}$ or $\ket{1}$, wait for a long delay (up to 15 s), and vary the readout scheme as detailed in Fig.~\ref{fig:leakage}.
Following the panel numbering of Fig.~\ref{fig:leakage}, measurement (a) provides us with a measure of the bright measurement error $\epsilon_b = 1.6(3) \times 10^{-2} \ \text{s}^{-1}$.
Subtracting the outcomes of measurements (a) and (c), or (b) and (d) provide two ways to measure the bit flip error $P_{0 \leftrightarrow 1}$.
Both approaches lead to statistically equal outcomes, with a nominal value close to 0, and an upper bound of $P_{0 \leftrightarrow 1} < 3.6 \times 10^{-3} \ \text{s}^{-1}$.
We have no independent measurements of the dark measurement error $\epsilon_d$ and the leakage error $P_{0,1 \rightarrow L}$.
We may therefore only extract $\epsilon_d+P_{0,1 \rightarrow L}$, which is done by subtracting measurements (b) and (d), or by subtracting the bit flip error from measurement (b).
Again, we have verified that these two approaches to measuring $\epsilon_d+P_{0 \rightarrow L} = 1.3(4) \times 10^{-2} \ \text{s}^{-1}$ and $\epsilon_d+P_{1 \rightarrow L} = 1.2(3) \times 10^{-2} \ \text{s}^{-1}$ provide statistically similar results.

\begin{figure*}[h]
    \centering
    \includegraphics[width=0.45\textwidth]{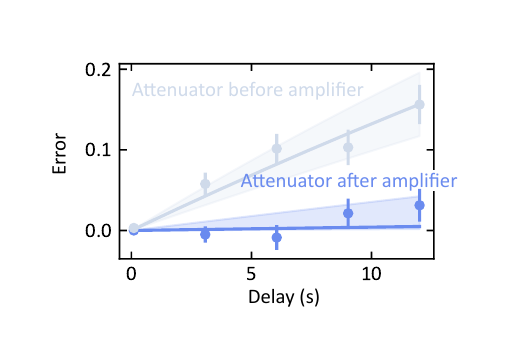}
    \caption{
    \textbf{Improvements in bit-flip errors.}
    By combining the measurements of Fig~\ref{fig:leakage}, we can extract the bit-flip error probability $P_{0 \leftrightarrow 1}$ as a function of delay time as described in Sec.~\ref{SI:leakage}.
    Here we show the resulting error when the attenuation was placed before or after the microwave amplifier (see Fig.~\ref{fig:mw_chain} for the microwave drive chain).
    }
    \label{fig:bitlip}
\end{figure*}

When attempting to combine these error sources following Eq.~(\ref{eq:epsilon_RB}), we exceed the total measured gate error by more than a factor 2.
The magnitude of these errors on the 10 s timescale is therefore not linearly related to their magnitude on the $<$ 1 s timescale.
We believe that the non-linear time evolution of these errors is most likely due to measurement errors $\epsilon_d$ or $\epsilon_b$ rather than leakage or bit-flip errors.
The development of a readout technique accounting for the sequence length could thus reduce the measured gate error.
These longer-timescale measurements can therefore not be used to quantitatively characterise the gate error, but are instead used here to illustrate the improvements made to the apparatus.
In Fig.~\ref{fig:leakage}(b), we show how improving laser extinction, through the use of double-pass acoustic-optic modulator setups, reduced the leakage error.
In Fig.~\ref{fig:bitlip}, we show how adding 22 dB attenuation after the gate microwave amplifier, effectively improving the microwave signal-to-noise ratio, reduced the probability of bit-flip errors.
This reduced the bit-flip error from being a dominant source of gate error ($1.9(5)\times 10^{-7}$ at a 13~$\upmu$s gate time) to a small contribution ($< 5 \times 10^{-9}$).

